\documentclass[12pt,a4paper]{panl}
\usepackage{cite}
\usepackage{wrapfig}
\usepackage{graphicx}
\usepackage{amssymb}
\usepackage{amsfonts}
\usepackage{amsmath}
\usepackage{longtable}
\usepackage{rotating}
\usepackage{lscape}
\usepackage{epsfig}
\usepackage{multirow}
\usepackage{caption}
\usepackage{subcaption}
\usepackage[utf8]{inputenc}
\usepackage[russian]{babel}

\originalTeX
\begin{document}

\title{Исследование эффективности измерения азимутальных   потоков в эксперименте MPD (NICA)  с фиксированной мишенью}
\maketitle
\authors{П.Е.\,Парфенов$^{a,b,c}$\footnote{E-mail: pparfenov@jinr.ru},
М.В.\,Мамаев$^{a,b,c}$,
А.В.\,Тараненко$^{a,c}$
}
\setcounter{footnote}{0}
\from{$^{a}$\,{\footnotesize Объединенный Институт Ядерных Исследований, Россия}}
\from{$^{b}$\,{\footnotesize Институт Ядерных Исследований Российской Академии Наук, Россия}}
\from{$^{c}$\,{\footnotesize Национальный Исследовательский Ядерный Университет <<МИФИ>>, Россия}}

\begin{abstract}
Исследование свойств сильно-взаимодействующей материи 
при больших относительных барионных плотностях является одной из ключевых научных задач эксперимента MPD ("Многоцелевой Детектор")
на ускорительном комплексе NICA. В работе исследуется эффективность измерения азимутальных коллективных потоков идентифицированных заряженных
адронов на установке MPD в режиме работы с фиксированной мишенью.

Studying the properties of strongly-interacting matter at high relative baryon densities is one of the key scientific goals of the MPD (Multi-Purpose Detector) experiment at the NICA accelerator complex. The performance of measuring the azimuthal collective flow of identified charged hadrons at the MPD facility in fixed-target mode is studied in this work.
\end{abstract}
\vspace*{6pt}


\label{sec:intro}
\vspace{-3mm}
\section*{Введение}
\vspace{-3mm}

Поиск сигналов начала деконфайнмента, фазового перехода первого рода и критической точки сильно-взаимодействующей ядерной (КХД)  материи является основой для программ
сканирования по энергии столкновения ядер от  $\sqrt{s_{NN}}$=2.4 до 11 ГэВ в системе центра масс \cite{Bzdak:2019pkr}. 
Изучения фазовой диаграммы КХД материи в области барионных плотностей превышающих нормальную ядерную материю в 3-10 раз
является одной из ключевых научных задач экспериментов:  ВM@N (“Барионная Материя на Нуклотроне”) и
MPD ("Многоцелевой Детектор") на ускорительном комплексе Nuclotron-NICA в ОИЯИ (г. Дубна)~\cite{Kekelidze:2018fvh,MPD:2022qhn}.\\
Одними из важнейших экспериментально наблюдаемых эффектов, которые чувствителны к свойствам КХД материи, образуемой  в релятивистских столкновениях тяжелых ионов,
являются азимутальные коллективные потоки в рождении адронов. Амплитуда азимутальных потоков 
определяется коэффициентами  $v_n=\left\langle \cos\left( n \left( \varphi - \Psi_n \right) \right) \right\rangle$   в разложении в ряд Фурье
зависимости выхода частиц от разницы между азимутальным углом импульса
частиц $\varphi$ и углом плоскости симметрии области перекрытия ядер $\Psi_n$, где $n$ -- порядок гармоники, и  скобки обозначают среднее значение по частицам
и событиям~\cite{Voloshin:2008dg}.  Благодаря своей чувствительности к  ранним временам столкновения, первые два
коэффициента $v_1$ (направленный поток) и $v_2$ (эллиптический поток) являются одними из самих чувствительных к ``жесткости''
уравнения состояния (EOS)  КХД материи~\cite{E895:2000maf, E895:1999ldn, HADES:2020lob, STAR:2013ayu, Sorensen:2023zkk}.
Новые высокоточные  дифференциальные измерения $v_n$ из эксперимента BM@N на Нуклотроне ($\sqrt{s_{NN}}$~= 2.4--3.5 ГэВ) и эксперимента MPD на
коллайдере NICA ($\sqrt{s_{NN}}$~= 4--11 ГэВ) помогут еще больше ограничить EOS симметричной материи в области высокой
барионной плотности ~\cite{Taranenko:2019uyv,Parfenov:2022brq}.\\
В настоящее время активно обсуждается возможность использования детектора MPD для изучения столкновений
тяжелых релятивистских ядер на фиксированной мишени (MPD-FXT). Это позволит: 1)
расширить диапазон энергий столкновения ядер, изучаемый в эксперименте MPD до $\sqrt{s_{NN}}$~= 2.4 ГэВ;
2) полностью решить проблему уменьшающейся частоты ядерных столкновений при уменьшении энергии столкновения, характерную для
работы ускорителя в коллайдерном режиме; 3) провести измерение столкновений одних и тех же ядер при одной и той же энергии в экспериментах BM@N и MPD.
Последнее обеспечит расширение динамического диапазона измерений, т.к. установки обладают различными аксептансами, а также проведение
сравнительных анализов для проверки и подтверждения получаемых результатов.
 В работе исследуется эффективность измерения азимутальных коллективных потоков идентифицированных заряженных
адронов на установке MPD-FXT.

\label{sec:MPDFXT}
\vspace{-3mm}
\section*{Детектор MPD-FXT в режиме работы с фиксированной мишенью}
\vspace{-2mm}

Для реализации программы MPD-FXT  в вакуумную трубу коллайдера NICA должна быть установлена мишень, представляющая из себя тонкую проволоку
диаметром 50-100 мкм, смещенную от центра канала на $\sim$ 1 см. Проволока должна быть установлена близко к торцу центрального барреля экспериментальной
установки MPD для эффективной регистрации частиц, рождающихся в столкновениях, см. Рис.~\ref{fig:MpdFxtLayout}. При этом установка  MPD будет состоять из 
тех же детекторных подсистем, что и в режиме коллайдера~\cite{MPD:2022qhn}: время-проекционная камера (TPC), времяпролетная камера (TOF),
электромагнитный калориметр (ECal), передний адронный калориметр (FHCal) и быстрый передний детектор (FFD).
%
\begin{figure}[thb!]
\begin{center}
\includegraphics[width=75mm]{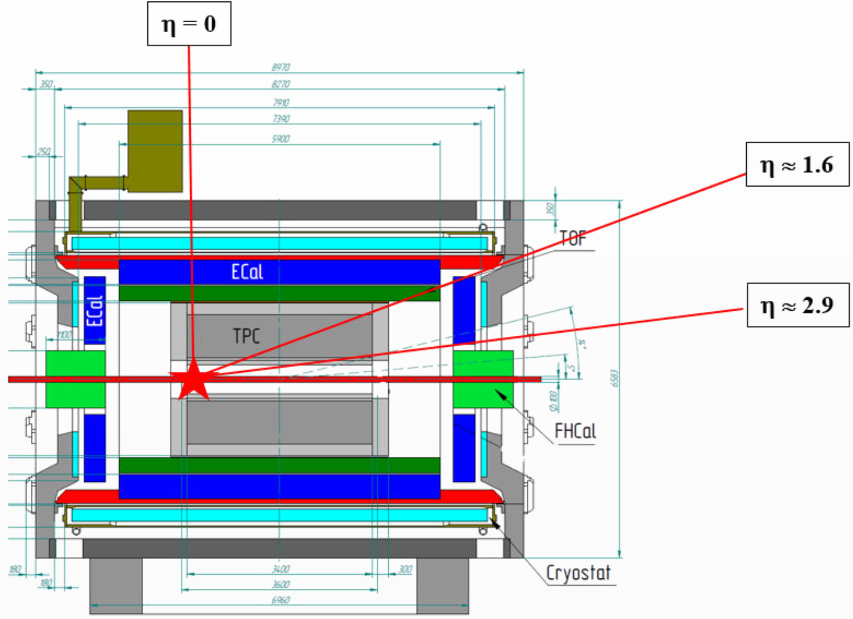}
\vspace{-3mm}
\caption{ Схема установки эксперимента MPD-FXT.}
\end{center}
\labelf{fig:MpdFxtLayout}
\vspace{-5mm}
\end{figure}
%
Идентификация заряженных частиц может быть выполнена с использованием ионизационых потерь (dE/dx)  в газовом объеме TPC,  
квадрата массы частицы ($m^{2}$) с использованием времени полета из TOF и импульса частицы из TPC. Плоскости симметрии могут бы оценены, используя
треки в TPC и выделения энергии в модулях правой части  FHCal.\\
Для  исследования  эффективности установки MPD-FXT для измерения направленного $v_1$  и эллиптического    $v_2$ потоков  протонов и заряженных пионов
было проведено численное моделирование 10М событий   Bi~+~Bi столкновений для следующих  энергий пучка $E_{kin}$~= 1.45, 2.92
и 4.65~АГэВ ($\sqrt{s_{NN}}$~= 2.5, 3 и 3.5~ГэВ). Для этого была использована модель UrQMD (версия 3.4) \cite{Bleicher:1999xi} с импульсно-зависимым
среднем полем.
На основе GEANT4 ~\cite{GEANT4:2002zbu} Монте-Карло моделирования отклика детекторных систем  MPD    и последующей реалистичной реконструкции сигналов
в MPDROOT, были получены события для анализа.

\label{sec:AnalysisDetails}
\vspace{-6mm}
\section*{Детали анализа}
\vspace{-3mm}

Для определения центральности столкновений использовалось распределение множественности ($N_{ch}$) реконструированных
треков заряженных частиц в TPC в диапазоне псевдобыстрот $0 < \eta < 2$. В качестве примера, открытыми символами
на левой части Рис.~\ref{fig:cent} показано распределение $N_{ch}$ для полностью реконструированных 
событий  Bi~+~Bi столкновений при энергии $E_{kin}$~= 1.45~АГэВ ($\sqrt{s_{NN}}$~= 2.5~ГэВ). 
Для оценки центральности и восстановления распределения прицельного параметра $b$ по множественности $N_{ch}$  образующихся заряженных частиц,
использовался ''$\Gamma$-fit'' метод, основанный на обратной теореме Байеса~\cite{Parfenov:2021ipw}. 
В данном подходе, распределение множественности $N_{ch}$  связано с  центральностью, определенной по прицельному параметру $c_b=\int\limits_{0}^{b}P(b')db'$,
через  флуктуационное ядро $P\left(N_{ch}|c_b\right)$:
\begin{eqnarray}
  P\left(N_{ch}|c_b\right) = \frac{1}{\Gamma\left(k\left(c_b\right)\right)\theta^{k\left(c_b\right)}} N_{ch}^{k\left(c_b\right)-1} e^{-\frac{N_{ch}}{\theta}},
\end{eqnarray}
где $\Gamma(k)$ гамма-функция с положительными параметрами $\theta=\frac{\sigma(N_{ch})}{\left\langle N_{ch} \right\rangle}$ и
$k = \frac{\left\langle N_{ch} \right\rangle}{\theta}$. Для параметризации средней множественности как функции $c_b$ можно использовать 
$\left\langle N_{ch} \right\rangle = N_{knee} \exp\left(\sum\limits_{j=1}^{3}a_jc_b^j\right)$,  с параметрами $a_j$, $N_{knee}$.
Функция распределения множественности  $P\left(N_{ch}\right)$ строится из  $P\left(N_{ch}|c_b\right)$ следующим
образом: $P\left(N_{ch}\right) = \int\limits_{0}^{1} P\left(N_{ch}|c_b\right) dc_b$. После нахождения оптимальных параметров ($a_j$, $N_{knee}$, $\theta$),
определяется условная вероятность $P\left(b|N_{ch}\right)$, используя обратную теорему Байеса: $P\left(b|N_{ch}\right) = P(b) \frac{P\left(N_{ch}|b\right)}{P\left(N_{ch}\right)}$.
%
\begin{figure}[thb!]
\begin{center}
  \centering
  \begin{subfigure}[b]{0.41\textwidth}
    \centering
    \includegraphics[width=\textwidth]{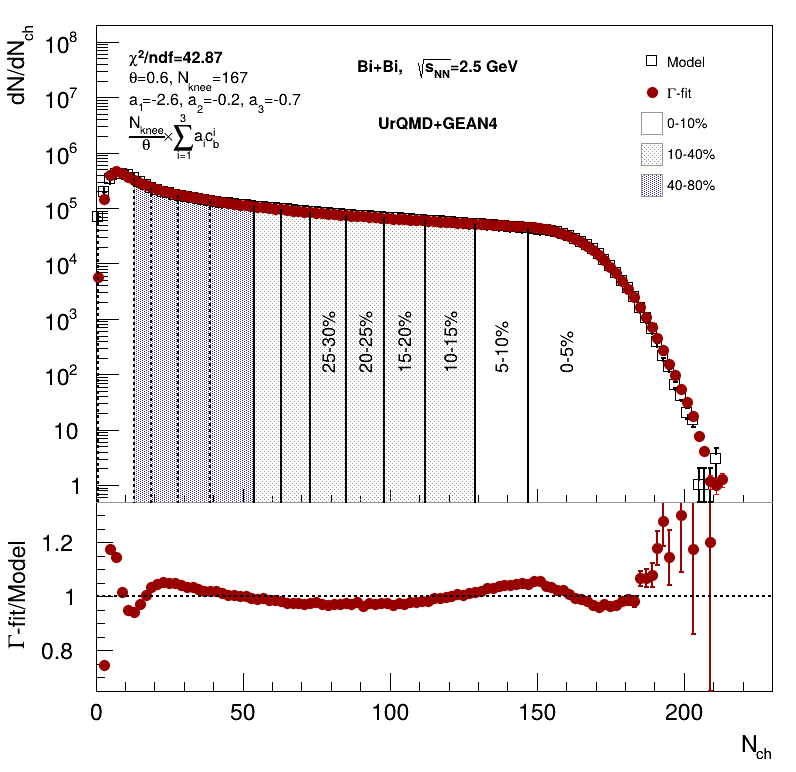}
  \end{subfigure}
  \hfill
  \begin{subfigure}[b]{0.41\textwidth}
    \centering
    \includegraphics[width=\textwidth]{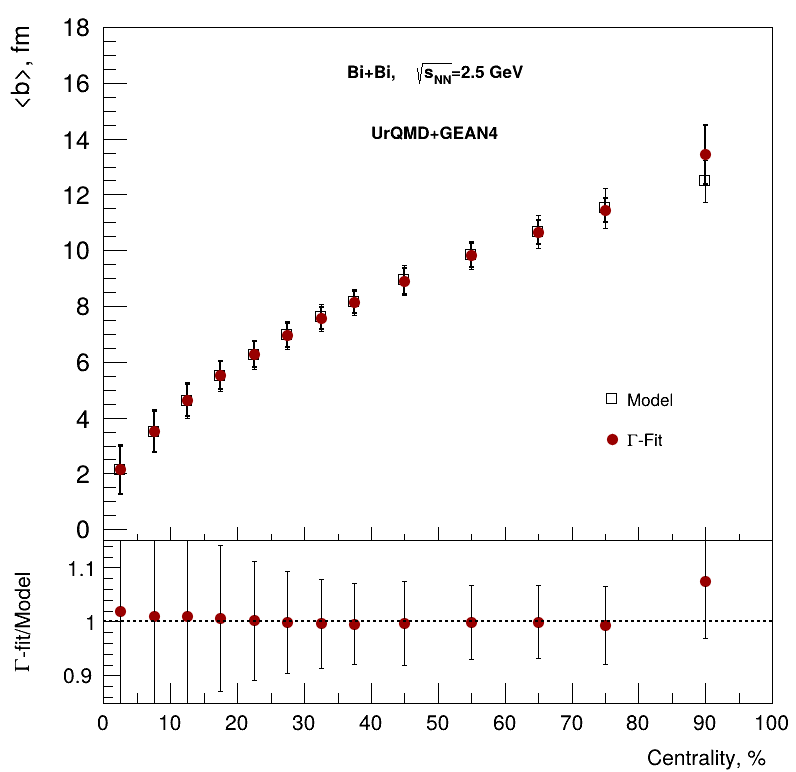}
  \end{subfigure}
  \hfill
  \vspace{-3mm}
  \caption{ Слева: Распределение множественности $N_{ch}$ заряженных частиц в  TPC (открытые символы) и
    результирующая функция параметризации $P\left(N_{ch}\right)$ (закрытые символы) для столкновений Bi~+~Bi при энергии $E_{kin}$~= 1.45~АГэВ ($\sqrt{s_{NN}}$~= 2.5~ГэВ).
      Внизу указано отношение параметризации и первоначального распределения множественности.
      Справа: Зависимость
      средних значений прицельного  параметра $\langle b \rangle$ от  центральности для модельных данных (открытые символы) и результата применения ``$\Gamma$-fit'' метода
      (закрытые символы). }
\end{center}
\labelf{fig:cent}
\vspace{-5mm}
\end{figure}
%
Закрытые символы на левой части Рис.~\ref{fig:cent} показывают полученную в результате подгонки параметров функцию параметризации $P\left(N_{ch}\right)$, а вертикальные линии
показывают полученные классы по центральности: 0-5\%, 5-10\%, ... и так далее. На правой части Рис.~\ref{fig:cent} показана зависимость
средних значений прицельного  параметра $\langle b \rangle$ от  центральности. Результаты, полученные из ``$\Gamma$-fit'' метода, основанного на обратной теореме Байеса, (закрытые символы) довольно хорошо сходятся с оценками непосредственно из модельных данных (открытые символы).\\
Для идентификации протонов и положительно заряженных пионов использовалась информация об ионизационых потерях (dE/dx) частиц в газовом объеме TPC и
времени пролета частиц в детекторе TOF.
Все частицы с отрицательным зарядом считались отрицательно заряженными пионами.
На левой  части Рис.~\ref{fig:pid} приведена зависимость $dE/dx$ от жесткости частиц $p/q$. Для описания использовалась параметризация эксперимента  ALICE (LHC),
использующая 5 параметров~\cite{Blum:2008nqe}:
\begin{eqnarray}
  f\left(\beta\gamma\right) = \frac{p_1}{\beta^{p_4}}\left[ p_2 - \beta^{p_4} - \ln\left( p_3 + \frac{1}{\left(\beta\gamma\right)^{p_5}} \right)  \right],
\end{eqnarray}
где $\beta^2 = p^2/(m^2 + p^2)$, $\beta\gamma = p/m$, а $p_i$ -- параметры аппроксимации.
После параметризации данной функцией были построены n-$\sigma$ распределения  ${\left(\left( dE/dx \right)^{measured} - f\left(\beta\gamma\right)\right)}/{f\left(\beta\gamma\right)}$
от жесткости $p/q$. Далее эти распределения были параметризованы функцией Гауса в разных диапазонах жесткости и
получены соответствующие величины средних квадратичных отклонений $\sigma\left( dE/dx \right)$.
Схожим образом были параметризованы  распределения квадратов масс ($m^{2}$) частиц в разных диапазонах жесткости и получены $\sigma\left( m^2 \right)$.
На правой части  Рис.~\ref{fig:pid} показана зависимость  $m^{2}$ частиц  от жесткости p/q.
Данные процедуры были проделаны для протонов и положительно заряженных пионов.
%
\begin{figure}[thb!]
  \begin{center}
    \centering
    \begin{subfigure}[b]{0.41\textwidth}
      \centering
      \includegraphics[width=\textwidth]{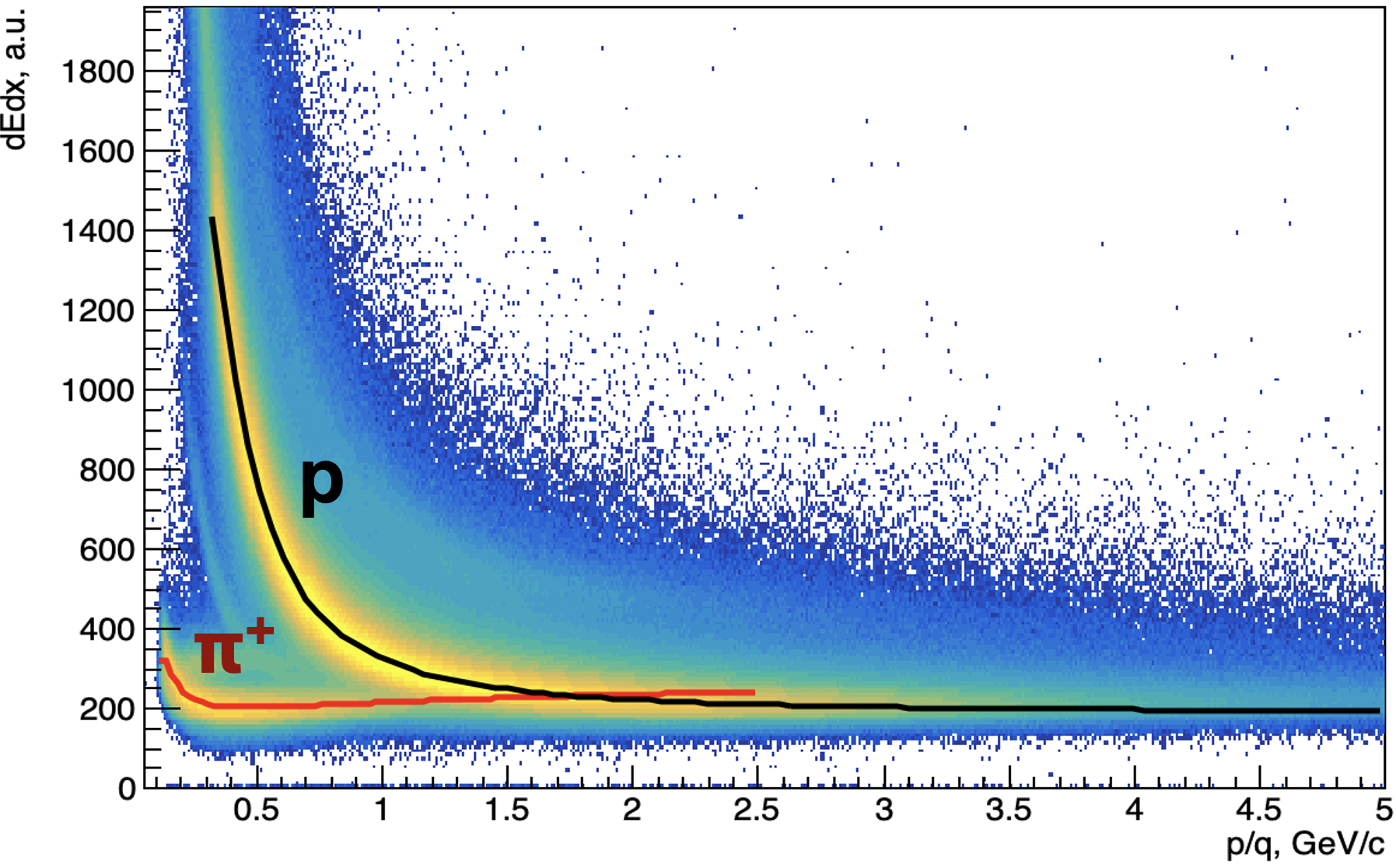}
    \end{subfigure}
    \hfill
    \begin{subfigure}[b]{0.42\textwidth}
      \centering
      \includegraphics[width=\textwidth]{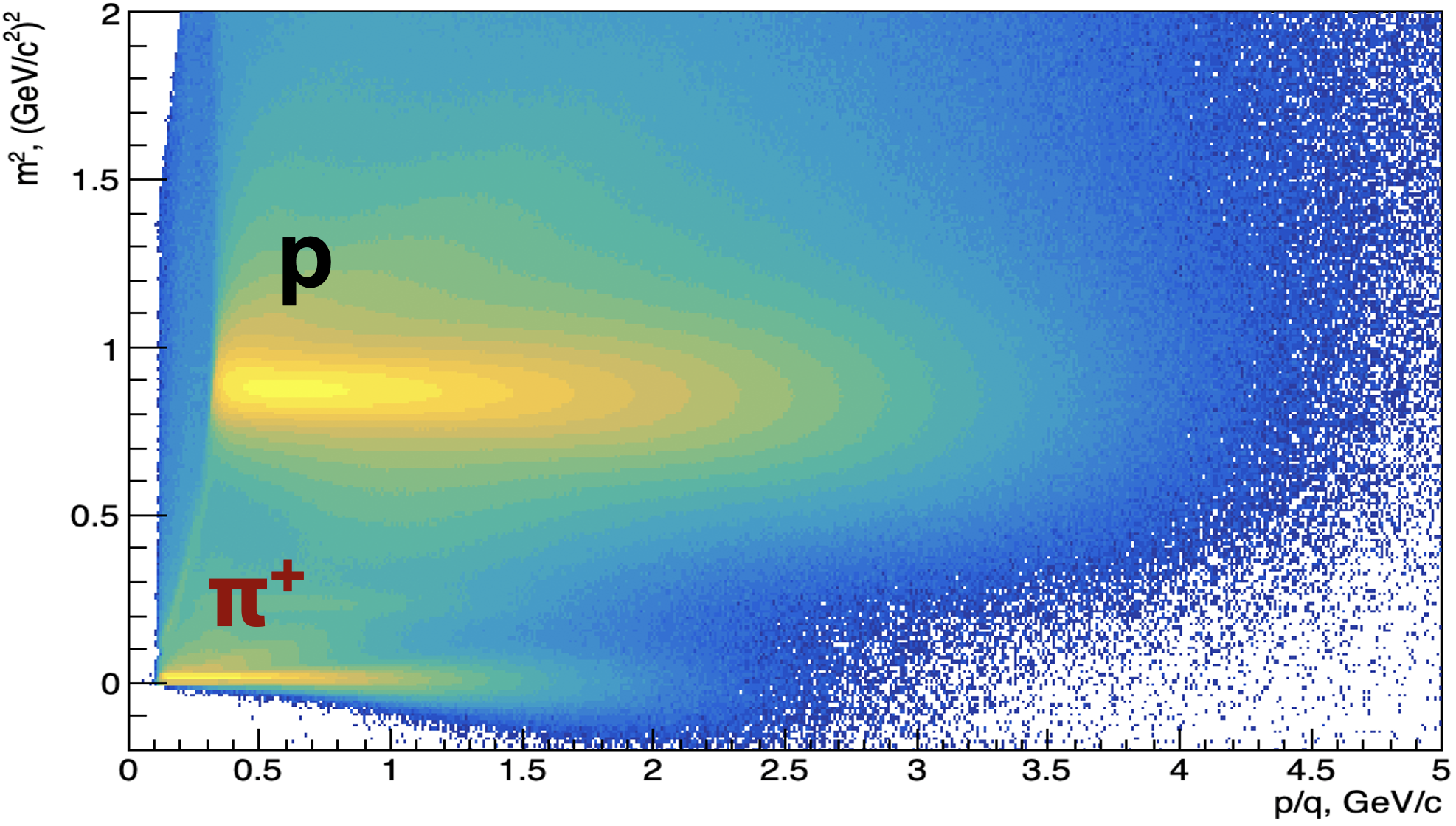}
    \end{subfigure}
    \hfill
    \vspace{-3mm}
    \caption{ Зависимость удельных потерь энергии  частиц в TPC $dE/dx$ (слева) и квадрата массы ($m^{2}$) частиц (справа) от жесткости  $p/q$ для столкновений Bi~+~Bi при энергии $E_{kin}$~= 1.45~АГэВ ($\sqrt{s_{NN}}$~= 2.5~ГэВ).}
  \end{center}
  \labelf{fig:pid}
  \vspace{-5mm}
  \end{figure}
%
Далее были определены новые координаты $(x,y)$ с помощью полученных параметризаций:
\begin{eqnarray}
  x_p = \frac{\left(\left( dE/dx \right)^{measured} - f_p\left(\beta\gamma\right)\right)}{f_p\left(\beta\gamma\right)\sigma_p\left( dE/dx \right)},\ y_p = \frac{m^2 - m_p^2}{\sigma_p\left( m^2 \right)},
\end{eqnarray}
где нижний индекс ''$p$'' означает тип частицы, для которой были определены координаты $(x,y)$, а $m_p^2$ -- табличное значение квадрата массы частицы.
В итоге, для идентификации протонов и $\pi^{+}$  использовались критерии отбора по координатам $(x,y)$: $\sqrt{x_p^2 + y_p^2}<2$, $\sqrt{x_\pi^2 + y_\pi^2}>3$ для протонов и $\sqrt{x_\pi^2 + y_\pi^2}<2$, $\sqrt{x_p^2 + y_p^2}>3$ для $\pi^{+}$.

Метод скалярного произведения был применен для получения  результатов для коэффициентов направленного $v_1$ и эллиптического $v_2$ потоков протонов и пионов.
В этом методе коэффициенты $v_n$ можно записать через вектора потока $Q_n$ и единичные вектора $u_n$~\cite{Selyuzhenkov:2007zi,Mamaev:2020lpi,Mamaev:2020qom,Mamaev:2023yhz}.
Единичный вектор $k$-й частицы в событии можно определить как $u_{n,k} = e^{in\varphi_k} = \left( \cos n\varphi_k, \sin n\varphi_k \right)$, где $\varphi_k$ -- азимутальный угол импульса частицы.
Вектор потока $Q_{n}$ определяется как взвешенная сумма всех единичных векторов $u_n$ в выбранной группе частиц в событии, называемом подсобытием: $Q_n = \sum_{k=0}^M \omega_k u_{n,k} = \left\vert Q_n \right\vert e^{in\Psi_n} $.
Здесь $M$ обозначает множественность частиц, входящих в данный вектор потока, а $\Psi_n$ -- угол плоскости симметрии.
Единичные вектора $u_n$ и вектора потока $Q_n$ также были взвешаны на эффективность при соответствующих значениях поперечного импульса $p_T$ и быстроты $y$ частицы.
На правой части Рис.~\ref{fig:eff} показана карта эффективности реконструкции протонов в плоскости $p_T$ от $y$. 
Величина эффективности определена как $\text{eff} = \frac{dN/dydp_T(reco)}{dN/dydp_T(sim)}$, где $dN/dydp_T(reco)$ -- число идентифицированных протонов в
соответствующем диапазоне $p_T$ и $y$, а $dN/dydp_T(sim)$ -- аналогичная величина, полученная из первоначальных модельных данных.
%
\begin{figure}[thb!]
  \begin{center}
    \centering
    \begin{subfigure}[b]{0.41\textwidth}
      \centering
      \includegraphics[width=\textwidth]{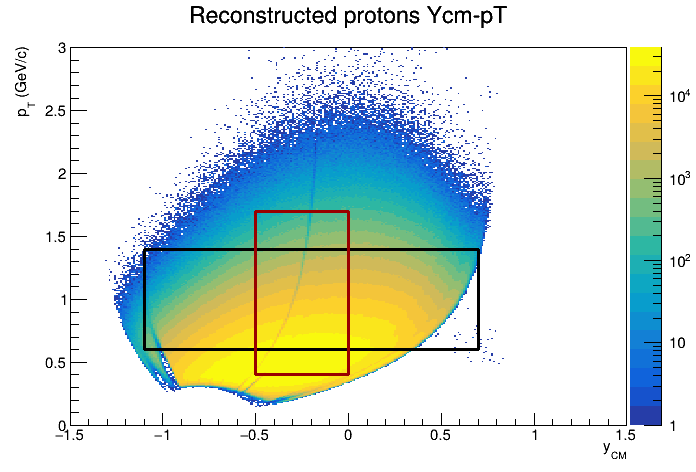}
    \end{subfigure}
    \hfill
    \begin{subfigure}[b]{0.41\textwidth}
      \centering
      \includegraphics[width=\textwidth]{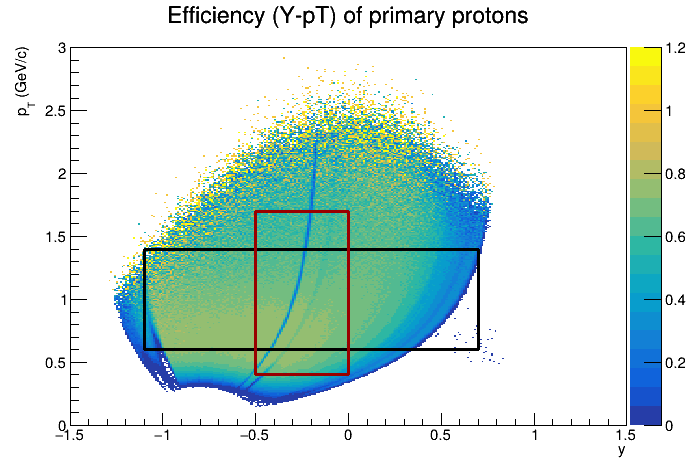}
    \end{subfigure}
    \hfill
    \vspace{-3mm}
    \caption{ Слева: Аксептанс протонов в плоскости $p_T$--$y$ для столкновений Bi~+~Bi при энергии $E_{kin}$~= 1.45~АГэВ ($\sqrt{s_{NN}}$~= 2.5~ГэВ).
     Справа: Карта эффективности реконструкции протонов в плоскости  $p_T$--$y$. Величина эффективности указана цветом.
     Рамками обозначены кинематические окна для зависимостей $v_n$ от $y$ (черные линии) и  $p_T$ (красные линии). }
  \end{center}
  \labelf{fig:eff}
  \vspace{-5mm}
  \end{figure}
%
Коэффициенты $v_n$ можно получить, проецируя вектор частицы $u_n$ на Q-вектор $Q_n$: $v_n = \left\langle u_n Q_n \right\rangle / R_n$, где $R_n$ называется поправочным
коэффициентом разрешения плоскости симметрии, а скобки обозначают среднее значение по частицам и событиям.
Аксептанс FHCal был разделен на  три подсобытия: F1, F2 и F3,  см. левую панель на Рис.~\ref{fig:flowSP}).
В дополнении было использовано  подсобытие (Tp) из TPC (см. правую панель на Рис.~\ref{fig:flowSP}) в кинематическом окне
$-1<y_{cm}<-0.6$, чтобы учесть возможную автокорреляцию между соседними подсобытиями (т.е. F1-F2 и F2-F3).
%
\begin{figure}[thb!]
  \begin{center}
    \centering
    \begin{subfigure}[b]{0.41\textwidth}
      \centering
      \includegraphics[width=\textwidth]{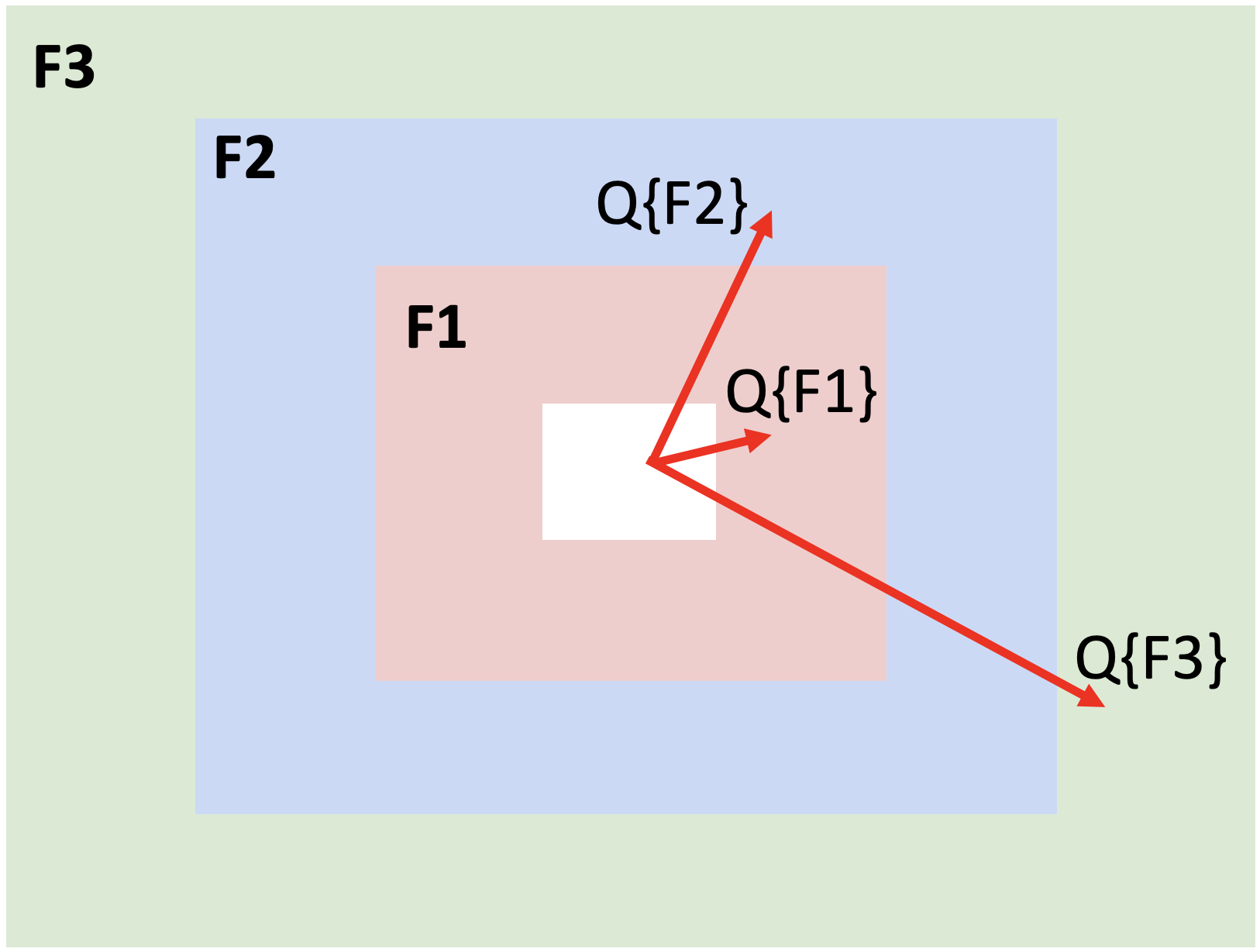}
    \end{subfigure}
    \hfill
    \begin{subfigure}[b]{0.42\textwidth}
      \centering
      \includegraphics[width=\textwidth]{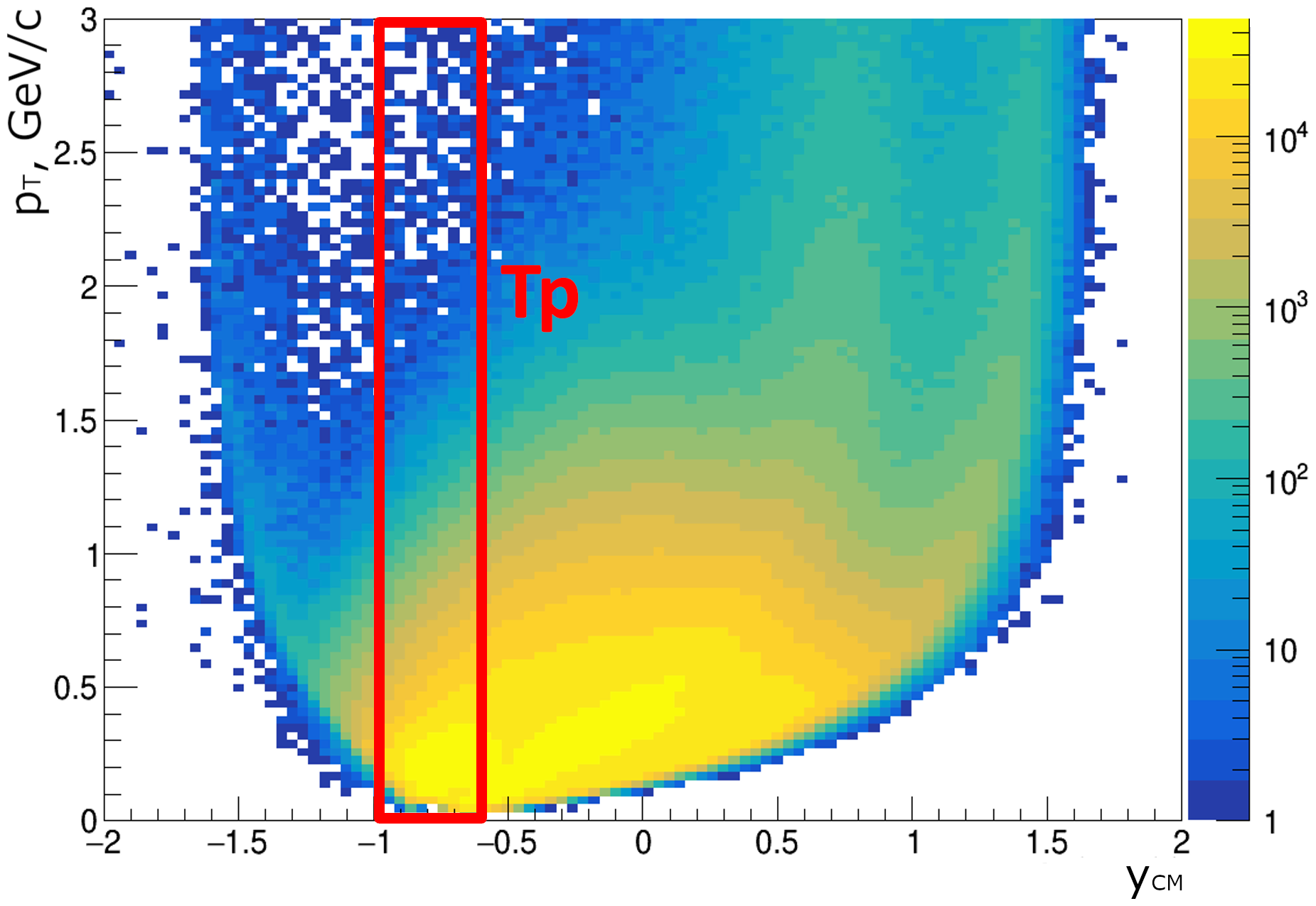}
    \end{subfigure}
    \hfill
    \vspace{-3mm}
    \caption{ Слева: Схематическое изображение модулей FHCal, разделенных на три подсобытия (F1, F2 и F3), обозначенных разными цветами.
      Стрелки обозначают Q-векторы каждого подсобытия. Справа: Схематическое изображение кинематического окна (в плоскости $p_T$-$y_{cm}$) для
      дополнительного подсобытия (Tp) из TPC. }
  \end{center}
  \labelf{fig:flowSP}
  \vspace{-5mm}
  \end{figure}
%
Для единичных векторов $u_n$ брались только треки частиц с количеством хитов в TPC  $N_{hit}>$22, чтобы гарантировать качество треков.
Отбор первичных треков был выполнен с помощью величины ближайшего расстояния трека до восстановленной первичной вершины: $|\text{DCA}|<1$~см.
Измерения $v_n$ были сделаны  относительно плоскости симметрии первого порядка ($n=1$):
\begin{eqnarray}
  v_1 = \frac{\left\langle u_1 Q_1^{F2} \right\rangle}{R_1\left\{F2\{Tp\}(F1,F3)\right\}},\ v_2 = \frac{\left\langle u_2 Q_1^{F1} Q_1^{F3} \right\rangle}{R_1\left\{F1(TpF3)\right\} R_1\left\{F3(TpF1)\right\}},
\end{eqnarray}
где поправочные коэффициенты разрешения плоскости симметрии $R_1$  рассчитывались следующим образом:
%
\begin{eqnarray}
  R_1\left\{F1(TpF3)\right\} = \sqrt{\frac{\left\langle Q_1^{F1} Q_1^{Tp} \right\rangle \left\langle Q_1^{F1} Q_1^{F3} \right\rangle}{\left\langle Q_1^{Tp} Q_1^{F3} \right\rangle}},\ 
  R_1\left\{F2\{Tp\}(F1,F3)\right\} = \frac{\left\langle Q_1^{F2} Q_1^{Tp} \right\rangle}{R_1\left\{Tp(F1F3)\right\}}.
\end{eqnarray}
%
Уравнения для $R_1\left\{F3(TpF1)\right\}$ и $R_1\left\{Tp(F1F3)\right\}$ эквивалентны уравнению для $R_1\left\{F1\left(TpF3 \right) \right\}$ с заменой подсобытий (F1,F3) и (F1,Tp) соответственно. Для коррекций результатов на неоднородности в азимутальном  аксептансе детектора, были использованы
поправки для Q-векторов: рецентеринг (recentering), поворот (twist) и масштабирование (rescale)~\cite{Selyuzhenkov:2007zi}.

\begin{figure}[thb!]
\begin{center}
  \centering
  \begin{subfigure}[b]{0.41\textwidth}
    \centering
    \includegraphics[width=\textwidth]{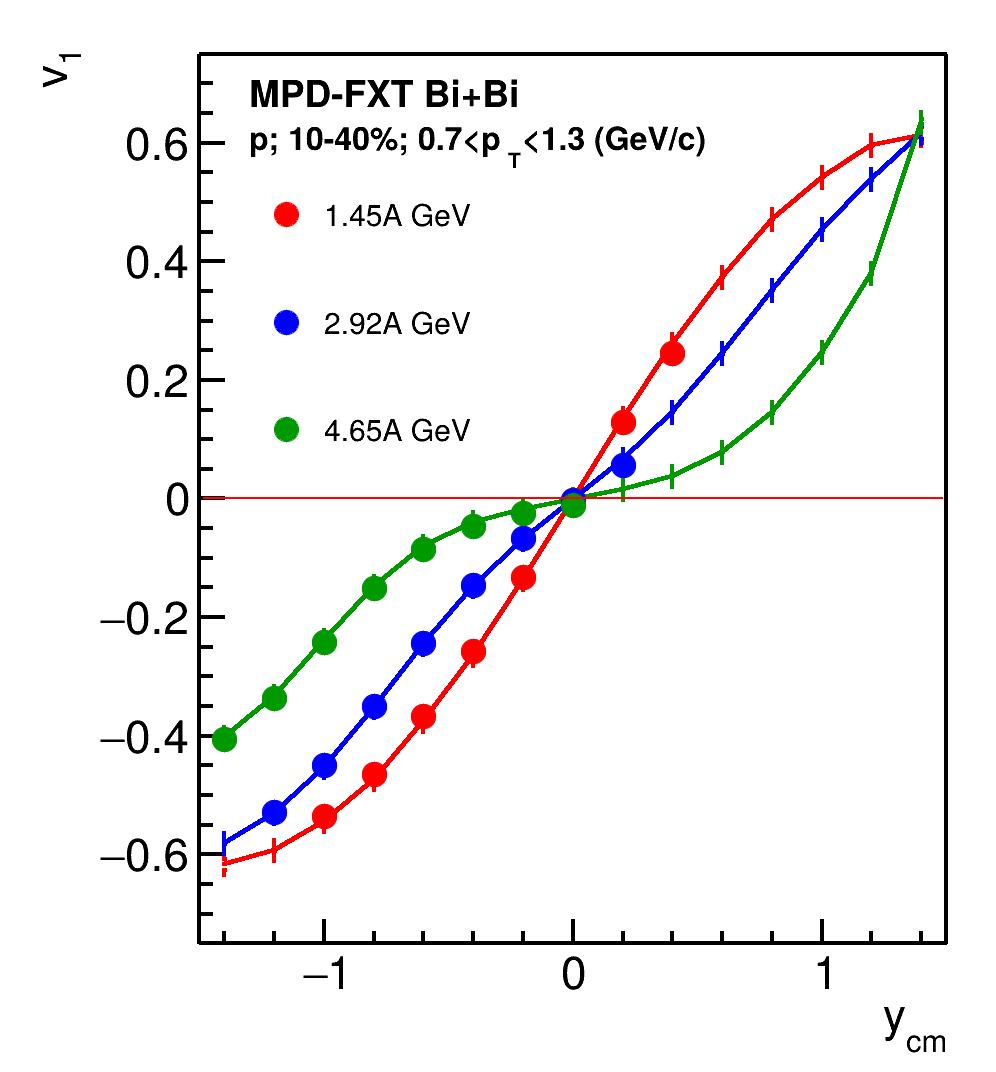}
  \end{subfigure}
  \hfill
  \begin{subfigure}[b]{0.41\textwidth}
    \centering
    \includegraphics[width=\textwidth]{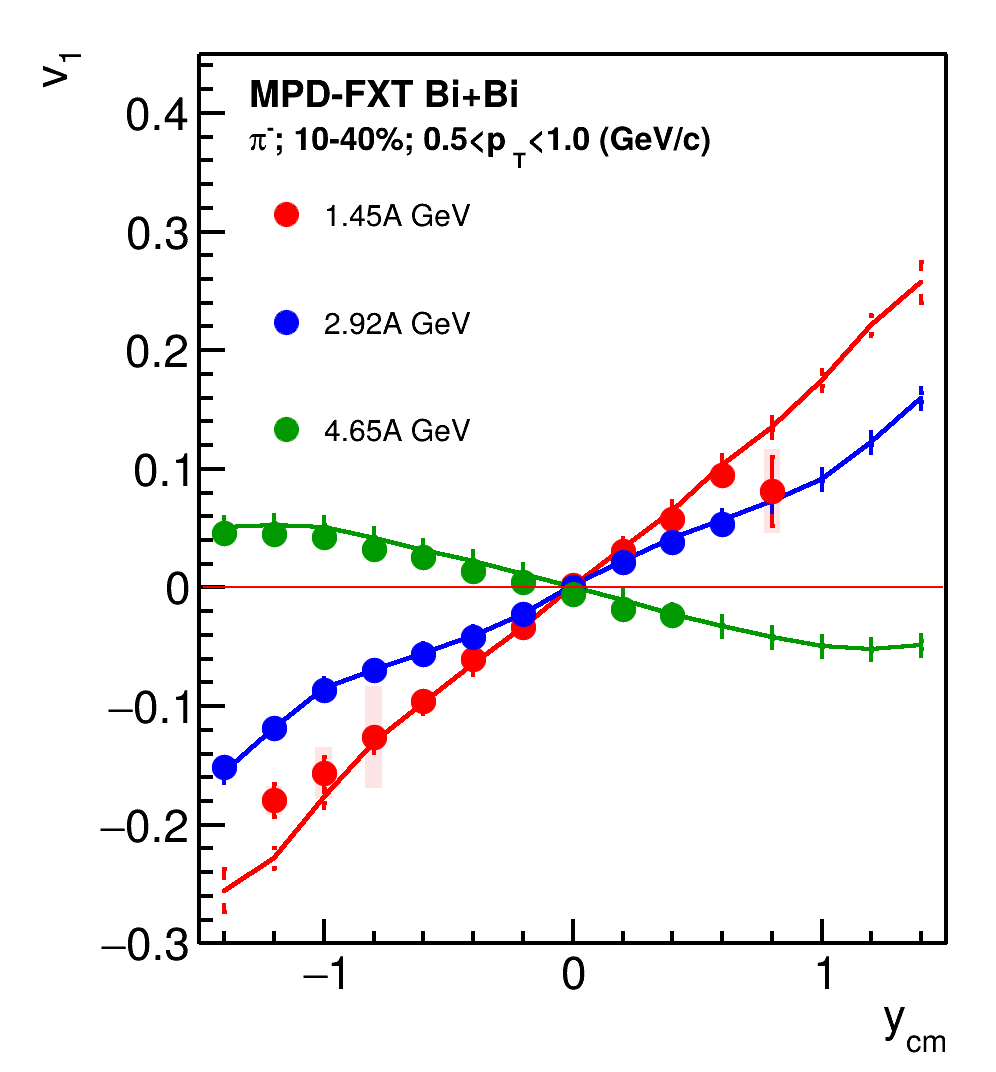}
  \end{subfigure}
  \hfill
  \vspace{-4mm}
  \caption{ Зависимость $v_1$ протонов (слева) и  пионов $\pi^{-}$ (справа) от быстроты $y_{cm}$   для 10-40\% центральных  Bi~+~Bi столкновений
    при $E_{kin}$~= 1.45, 2.92 и 4.65 АГэВ.
   Закрытыми  символами показаны значения $v_1^{reco}$ из анализа полностью реконструированных событий, а сплошными линиями значения $v_1^{true}$ напрямую из модели.}
\end{center}
\labelf{fig:v1bes}
\vspace{-6mm}
\end{figure}

\label{sec:results}
\vspace{-6mm}
\section*{Результаты}
\vspace{-3mm}

На рисунках~\ref{fig:v1bes}, \ref{fig:v2bes} показаны зависимости $v_1(y_{cm})$ и  $v_2(p_T)$ потоков протонов и  пионов $\pi^{-}$ для 10-40\% центральных  Bi~+~Bi столкновений
при $E_{kin}$~= 1.45, 2.92 и 4.65 АГэВ ($\sqrt{s_{NN}}$~= 2.5, 3 и 3.5~ГэВ, соответственно). Закрытые символы представляют результаты $v_n^{reco}$ анализа
полностью реконструированных данных, а сплошные линии показывают $v_n^{true}$ результаты, полученные напрямую из модели UrQMD. Разные цвета представляют разные энергии.
%
\begin{figure}[thb!]
\begin{center}
  \centering
  \begin{subfigure}[b]{0.41\textwidth}
    \centering
    \includegraphics[width=\textwidth]{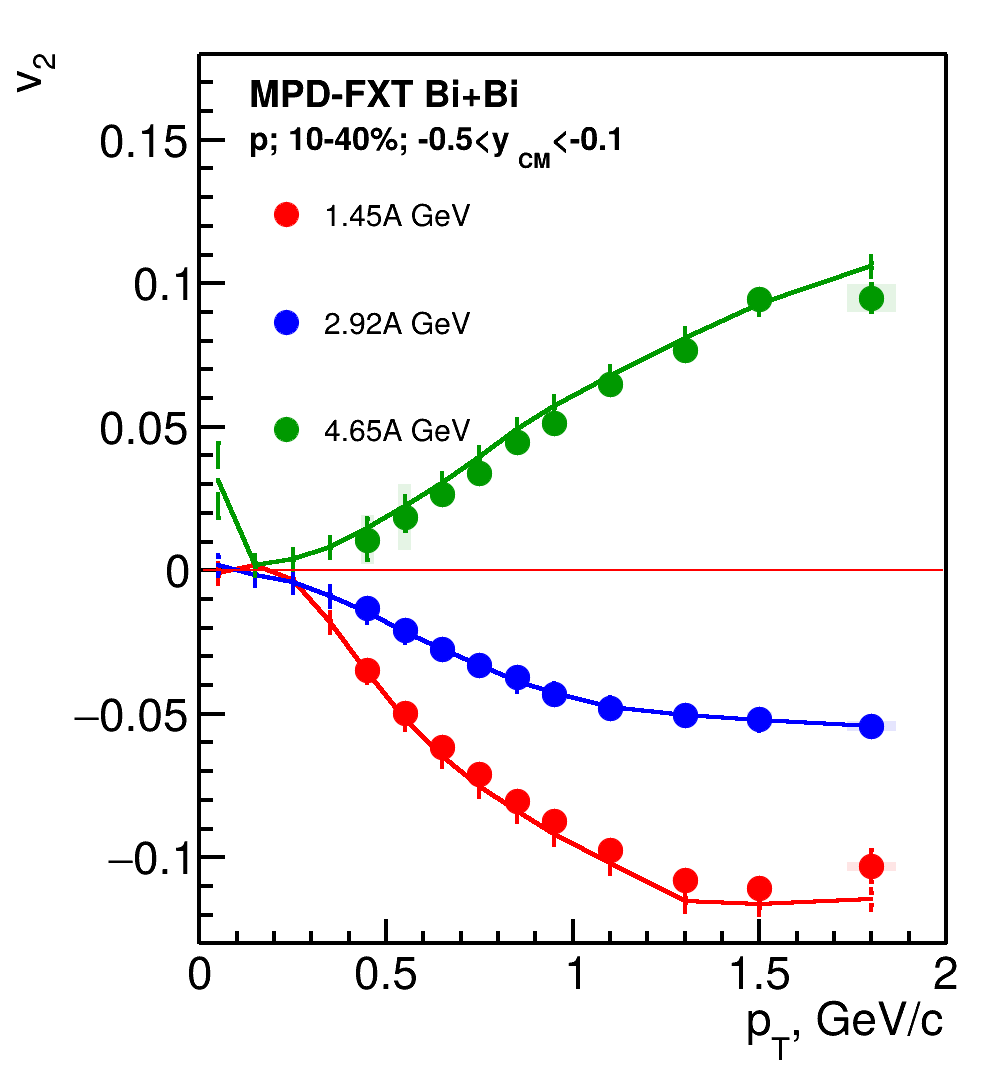}
  \end{subfigure}
  \hfill
  \begin{subfigure}[b]{0.41\textwidth}
    \centering
    \includegraphics[width=\textwidth]{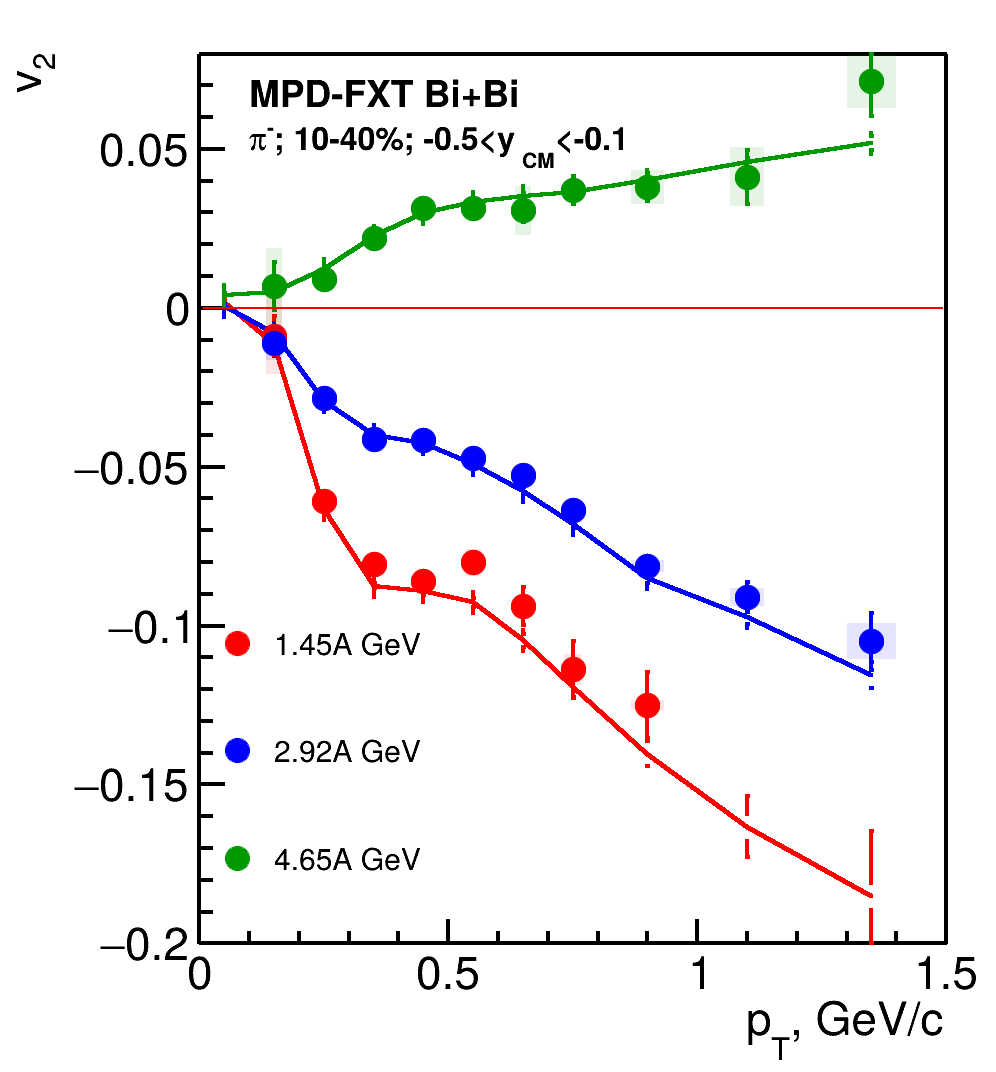}
  \end{subfigure}
  \hfill
  \vspace{-3mm}
  \caption{Зависимость $v_2$ протонов (слева) и  пионов $\pi^{-}$ (справа) от $p_{T}$   для 10-40\% центральных  Bi~+~Bi столкновений
    при $E_{kin}$~= 1.45, 2.92 и 4.65 АГэВ.
   Закрытыми  символами показаны значения $v_2^{reco}$ из анализа полностью реконструированных событий, а сплошными линиями значения $v_2^{true}$ напрямую из модели.}
\end{center}
\labelf{fig:v2bes}
\vspace{-6mm}
\end{figure}
%
Сравнение показывает, что результаты $v_n^{reco}$, полученные из анализа реконструированных данных и из модели UrQMD $v_n^{true}$  хорошо согласуются в области быстрот
$y_{cm} \lessapprox$ 0.5. Расхождения в передней области  быстрот являются результатом плохого акспетанса и эффективности идентификации частиц при $y_{cm} > $ 0.5 в MPD-FXT.

\begin{figure}[thb!]
  \begin{center}
  \includegraphics[width=65mm]{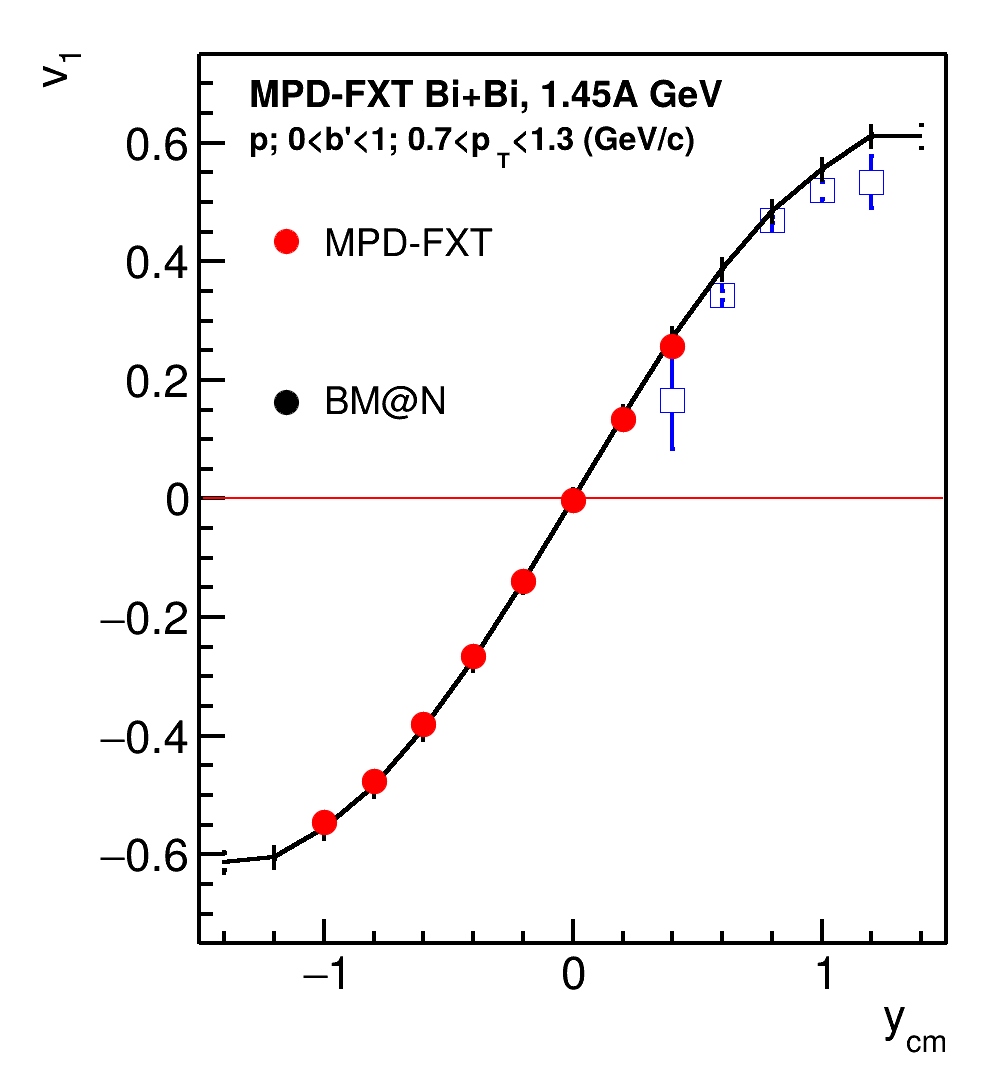}
  \vspace{-3mm}
  \caption{ Зависимость $v_1$ протонов от быстроты  $y_{cm}$ в столкновениях Bi~+~Bi при $E_{kin}$~= 1.45 АГэВ, символами показаны результаты анализа
    полностью реконструированных событий моделирования  экспериментов MPD-FXT и BM@N. }
  \end{center}
  \labelf{fig:mpdbmn}
  \vspace{-5mm}
  \end{figure}
%
Результаты анализа $v_1(y_{cm})$ протонов также сравнивались с результатами исследования эффективности измерения потоков в эксперименте BM@N~\cite{Mamaev:2023yhz} для 
Bi~+~Bi столкновений при $E_{kin}$~= 1.45 АГэВ, см. Рис.~\ref{fig:mpdbmn}.
Сплошные линии представляют $v_1^{true}$ результаты из модели UrQMD, а  символы -- результаты анализа полностью реконструированных событий MPD-FXT и BM@N.
Сравнение показывает, что MPD-FXT лучше сможет проводить измерения потоков в области задних и средних быстрот, в то время как
BM@N лучше справляется с точностью измерений в области передних быстрот.
В этом отношении эксперименты MPD-FXT и BM@N дополняют друг друга и обеспечивают более широкий охват области быстрот для дифференциальных измерений потоков.

\label{sec:summary}
\vspace{-6mm}
\section*{Заключение}
\vspace{-3mm}

В настоящей работе изучалась эффективность измерений направленного ($v_1$) и эллиптического ($v_2$) потоков протонов и заряженных
пионов на установке MPD в режиме работы с фиксированной мишенью MPD-FXT. Было проведено моделирование эксперимента  MPD-FXT
с большой выборкой событий модели UrQMD, для Bi~+~Bi столкновений при энергиях  $\sqrt{s_{NN }}$~= 2.5, 3 и 3.5~ГэВ,  в качестве входных данных. 
Для достижения реалистичной работы детекторных подсистем MPD-FXT использовался транспортный код GEANT4 и процедуры реконструкции из MPDROOT.
Были протестированы реалистичные процедуры оценки центральности столкновений,   плоскости симметрии, а также
методы расчета поправочного коэффициента разрешения плоскости симметрии. 
Полученные результаты эксперимента MPD-FXT были проверены для измерений $v_1$ и $v_2$  протонов и заряженных пионов в зависимости от быстроты ($y_{cm}$)
и поперечного импульса ($p_{T}$). Детальное сравнение результатов, полученных при анализе полностью реконструированных данных и
данных на уровне генератора UrQMD, позволило сделать вывод, что система MPD-FXT
будет достаточной для детальных дифференциальных измерений направленного и эллиптического потока в области быстрот $y_{см} \lessapprox$ 0.5.
Сделан вывод, что эксперименты MPD-FXT и BM@N могут хорошо
дополнять друг друга и обеспечивать более широкий охват по быстроте для дифференциальных измерений потоков.

\vspace{-6mm}
\section*{Благодарности}
\vspace{-3mm}

Работа поддержана Министерством науки и высшего образования РФ, проект
"Фундаментальные и прикладные исследования на экспериментальном комплексе класса мегасайенс NICA" FSWU-2024-0024

\vspace{-4mm}
\bibliographystyle{pepan}
{\bibliography{pepan_biblio}}

\begin{thebibliography}{10}
\def\selectlanguageifdefined#1{
\expandafter\ifx\csname date#1\endcsname\relax
\else\selectlanguage{#1}\fi}
\providecommand*{\href}[2]{{\small #2}}
\providecommand*{\url}[1]{{\small #1}}
\providecommand*{\BibUrl}[1]{\url{#1}}
\providecommand{\BibAnnote}[1]{}
\providecommand*{\BibEmph}[1]{\emph{#1}}
\ProvideTextCommandDefault{\cyrdash}{\hbox to.8em{--\hss--}}
\providecommand*{\BibDash}{}

\bibitem{Bzdak:2019pkr}
\selectlanguageifdefined{english}
\BibEmph{Bzdak A., Esumi S., Koch V., Liao J., Stephanov M., Xu N.} {Mapping
  the Phases of Quantum Chromodynamics with Beam Energy Scan}~//
  \href{http://dx.doi.org/10.1016/j.physrep.2020.01.005}{Phys. Rept.} \BibDash
\newblock 2020. \BibDash
\newblock V. 853. \BibDash
\newblock P.~1--87. \BibDash
\newblock arXiv:1906.00936.

\bibitem{Kekelidze:2018fvh}
\selectlanguageifdefined{english}
\BibEmph{Kekelidze V.D.} {Heavy Ion Collisions: Baryon Density Frontier}~//
  \href{http://dx.doi.org/10.1134/S1063779618040330}{Phys. Part. Nucl.}
  \BibDash
\newblock 2018. \BibDash
\newblock V.~49, no.~4. \BibDash
\newblock P.~457--472.

\bibitem{MPD:2022qhn}
\selectlanguageifdefined{english}
\BibEmph{Abgaryan V. et~al.} [MPD Collaboration] {Status and initial physics
  performance studies of the MPD experiment at NICA}~//
  \href{http://dx.doi.org/10.1140/epja/s10050-022-00750-6}{Eur. Phys. J. A}.
  \BibDash
\newblock 2022. \BibDash
\newblock V.~58, no.~7. \BibDash
\newblock P.~140.

\bibitem{Voloshin:2008dg}
\selectlanguageifdefined{english}
\BibEmph{Voloshin S.A., Poskanzer A.M., Snellings R.} {Collective phenomena in
  non-central nuclear collisions}~//
  \href{http://dx.doi.org/10.1007/978-3-642-01539-7_10}{Landolt-Bornstein}.
  \BibDash
\newblock 2010. \BibDash
\newblock V.~23. \BibDash
\newblock P.~293.

\bibitem{E895:2000maf}
\selectlanguageifdefined{english}
\BibEmph{Liu H. et~al.} [E895 Collaboration] {Sideward flow in Au + Au
  collisions between 2-A-GeV and 8-A-GeV}~//
  \href{http://dx.doi.org/10.1103/PhysRevLett.84.5488}{Phys. Rev. Lett.}
  \BibDash
\newblock 2000. \BibDash
\newblock V.~84. \BibDash
\newblock P.~5488--5492.

\bibitem{E895:1999ldn}
\selectlanguageifdefined{english}
\BibEmph{Pinkenburg C. et~al.} [E895 Collaboration] {Elliptic flow: Transition
  from out-of-plane to in-plane emission in Au + Au collisions}~//
  \href{http://dx.doi.org/10.1103/PhysRevLett.83.1295}{Phys. Rev. Lett.}
  \BibDash
\newblock 1999. \BibDash
\newblock V.~83. \BibDash
\newblock P.~1295--1298.

\bibitem{HADES:2020lob}
\selectlanguageifdefined{english}
\BibEmph{Adamczewski-Musch J. et~al.} [HADES Collaboration] {Directed,
  Elliptic, and Higher Order Flow Harmonics of Protons, Deuterons, and Tritons
  in $\mathrm{Au}+\mathrm{Au}$ Collisions at $\sqrt{{s}_{NN}}=2.4\text{ }\text{
  }\mathrm{GeV}$}~//
  \href{http://dx.doi.org/10.1103/PhysRevLett.125.262301}{Phys. Rev. Lett.}
  \BibDash
\newblock 2020. \BibDash
\newblock V. 125. \BibDash
\newblock P.~262301.

\bibitem{STAR:2013ayu}
\selectlanguageifdefined{english}
\BibEmph{Adamczyk L. et~al.} [STAR Collaboration] {Elliptic flow of identified
  hadrons in Au+Au collisions at $\sqrt{s_{NN}}=$ 7.7-62.4 GeV}~//
  \href{http://dx.doi.org/10.1103/PhysRevC.88.014902}{Phys. Rev. C}. \BibDash
\newblock 2013. \BibDash
\newblock V.~88. \BibDash
\newblock P.~014902.

\bibitem{Sorensen:2023zkk}
\selectlanguageifdefined{english}
\BibEmph{Sorensen A., others.} {Dense nuclear matter equation of state from
  heavy-ion collisions}~//
  \href{http://dx.doi.org/10.1016/j.ppnp.2023.104080}{Prog. Part. Nucl. Phys.}
  \BibDash
\newblock 2024. \BibDash
\newblock V. 134. \BibDash
\newblock P.~104080. \BibDash
\newblock arXiv:2301.13253.

\bibitem{Taranenko:2019uyv}
\selectlanguageifdefined{english}
\BibEmph{Taranenko A.} {Anisotropic flow measurements from RHIC to SIS}~//
  \href{http://dx.doi.org/10.1051/epjconf/201920403009}{EPJ Web Conf.} \BibDash
\newblock 2019. \BibDash
\newblock V. 204. \BibDash
\newblock P.~03009.

\bibitem{Parfenov:2022brq}
\selectlanguageifdefined{english}
\BibEmph{Parfenov P.} {Model Study of the Energy Dependence of Anisotropic Flow
  in Heavy-Ion Collisions at $\sqrt {s_{NN}}$ = 2\textendash{}4.5 GeV}~//
  \href{http://dx.doi.org/10.3390/particles5040040}{Particles}. \BibDash
\newblock 2022. \BibDash
\newblock V.~5, no.~4. \BibDash
\newblock P.~561--579.

\bibitem{Bleicher:1999xi}
\selectlanguageifdefined{english}
\BibEmph{Bleicher M., others.} {Relativistic hadron hadron collisions in the
  ultrarelativistic quantum molecular dynamics model}~//
  \href{http://dx.doi.org/10.1088/0954-3899/25/9/308}{J. Phys. G}. \BibDash
\newblock 1999. \BibDash
\newblock V.~25. \BibDash
\newblock P.~1859--1896.

\bibitem{GEANT4:2002zbu}
\selectlanguageifdefined{english}
\BibEmph{Agostinelli S. et~al.} [GEANT4 Collaboration] {GEANT4--a simulation
  toolkit}~// \href{http://dx.doi.org/10.1016/S0168-9002(03)01368-8}{Nucl.
  Instrum. Meth. A}. \BibDash
\newblock 2003. \BibDash
\newblock V. 506. \BibDash
\newblock P.~250--303.

\bibitem{Parfenov:2021ipw}
\selectlanguageifdefined{english}
\BibEmph{Parfenov P., Idrisov D., Luong V.B., Taranenko A.} {Relating Charged
  Particle Multiplicity to Impact Parameter in Heavy-Ion Collisions at NICA
  Energies}~// \href{http://dx.doi.org/10.3390/particles4020024}{Particles}.
  \BibDash
\newblock 2021. \BibDash
\newblock V.~4, no.~2. \BibDash
\newblock P.~275--287.

\bibitem{Blum:2008nqe}
\selectlanguageifdefined{english}
\BibEmph{Blum W., Rolandi L., Riegler W.}
  \href{http://dx.doi.org/10.1007/978-3-540-76684-1}{{Particle detection with
  drift chambers}} Particle Acceleration and Detection. \BibDash
\newblock 2008. \BibDash
\newblock ISBN:~\href{http://isbndb.com/search-all.html?kw=978-3-540-76683-4,
  978-3-540-76684-1}{978-3-540-76683-4, 978-3-540-76684-1}.

\bibitem{Selyuzhenkov:2007zi}
\selectlanguageifdefined{english}
\BibEmph{Selyuzhenkov I., Voloshin S.} {Effects of non-uniform acceptance in
  anisotropic flow measurement}~//
  \href{http://dx.doi.org/10.1103/PhysRevC.77.034904}{Phys. Rev. C}. \BibDash
\newblock 2008. \BibDash
\newblock V.~77. \BibDash
\newblock P.~034904.

\bibitem{Mamaev:2020lpi}
\selectlanguageifdefined{english}
\BibEmph{Mamaev M. et~al.} [HADES Collaboration] {Directed flow of protons with
  the event plane and scalar product methods in the HADES experiment at
  SIS18}~// \href{http://dx.doi.org/10.1088/1742-6596/1690/1/012122}{J. Phys.
  Conf. Ser.} \BibDash
\newblock 2020. \BibDash
\newblock V. 1690, no.~1. \BibDash
\newblock P.~012122.

\bibitem{Mamaev:2020qom}
\selectlanguageifdefined{english}
\BibEmph{Mamaev M. et~al.} [HADES Collaboration] {Estimating Non-Flow Effects
  in Measurements of Anisotropic Flow of Protons with the HADES Experiment at
  GSI}~// \href{http://dx.doi.org/10.1134/S1063779622020514}{Phys. Part. Nucl.}
  \BibDash
\newblock 2022. \BibDash
\newblock V.~53, no.~2. \BibDash
\newblock P.~277--281.

\bibitem{Mamaev:2023yhz}
\selectlanguageifdefined{english}
\BibEmph{Mamaev M., Taranenko A.} {Toward the System Size Dependence of
  Anisotropic Flow in Heavy-Ion Collisions at $\sqrt {s_{NN}}$= 2\textendash{}5
  GeV}~// \href{http://dx.doi.org/10.3390/particles6020036}{Particles}.
  \BibDash
\newblock 2023. \BibDash
\newblock V.~6, no.~2. \BibDash
\newblock P.~622--637.

\end{thebibliography}

\end{document}